# The effect of surface transport on water desalination by porous electrodes undergoing capacitive charging

Amit N. Shocron,[1] Matthew E. Suss[1]

1. Faculty of Mechanical Engineering, Technion – Israel Institute of Technology, Haifa, Israel

**Abstract**

Capacitive deionization (CDI) is a technology in which water is desalinated by ion electrosorption into the electric double layers (EDLs) of charging porous electrodes. In recent years significant advances have been made in modeling the charge and salt dynamics in a CDI cell, but the possible effect of surface transport within diffuse EDLs on these dynamics has not been investigated. We here present theory which includes surface transport in describing the dynamics of a charging CDI cell. Through our numerical solution to the presented models, the possible effect of surface transport on the CDI process is elucidated. While at some model conditions surface transport enhances the rate of CDI cell charging, counter-intuitively this additional transport pathway is found to slow down cell charging at other model conditions.

**Introduction**

Capacitive deionization (CDI) is an emerging technology typically applied to the desalination of brackish feedwaters[1], but also to organic solvent remediation[2], water softening[3], ion separations[4], microfluidic sample preparation[5], and sea water desalination[6]. The classical CDI cell consists of two porous carbon electrodes sandwiching a porous dielectric separator, see Figure 1a. The electrodes are charged by either a constant voltage or constant current[7], resulting in the formation of electric double layers (EDLs) in the micropores of the electrodes and the electrosorption of salt ions[8]. The feedwater flows most often between the two charging electrodes[1], although flow can also be through the electrodes themselves[9–11]. Alternative CDI cell architectures utilize suspension electrodes, such as flow or fluidized bed electrodes[6,12,13], and include ion exchange membranes along the inner electrode surfaces[14]. The porous electrodes of CDI cells typically contain a multiscale pore structure, which includes a through-electrode network of macropores which access smaller micropores[15]. The layout of the electrode's pore structure can vary for different materials, as for example, activated carbon electrodes often consist of a collection of bound micron-scale microporous carbon particles[16], while hierarchical carbon aerogel monoliths (HCAMs) have micropores etched into the walls of a macroporous monolith via thermal activation[9,17].

Surface transport refers the movement of ions in EDLs tangentially to the charged interface due to, for example, tangential electric fields or tangential gradients in ion concentration.[18] There is an extensive literature investigating the effects of surface transport in electrokinetic systems employing dielectric media such as planar dielectric walls[19–23], porous dielectric media[24–28], dielectric solid-liquid colloidal suspensions[29–31]. and also for electrochemical systems employing charging planar metal electrodes[32,33]. By contrast, the literature is far sparser on the topic of surface transport in charging conductive porous media such as porous carbons electrodes, despite the widespread application of such electrodes in energy storage systems and in water desalination by CDI[15,34]. To our knowledge, the latter literature consists solely of the work of Mirzadeh et al., who presented theory and a numerical model capturing the effect of surface transport on the charging dynamics of the porous electrodes of supercapacitors[35], and of Vol'fkovich et al. who presented measurements of surface conductivity as a function of applied potential in porous carbons[36]. We know of no previous works which investigate the effect of surface transport in porous electrodes on desalination and charging dynamics in a CDI cell. Surface transport (and more specifically surface conductivity) may be significant in CDI, as during cell charging the electrosorption of ions into the EDL from the bulk electrolyte can result in a high surface to bulk conductivity ratio (a high Dukhin number)[37]. We here develop the theory of surface transport in CDI cells, and apply the theory to understand the effect of surface transport on desalination and cell charging dynamics in both uni-and multi-scale porous electrodes. Counter-intuitively, we find that at some model conditions, the effect of surface transport (an additional transport pathway for ions) is to slow down cell charging.

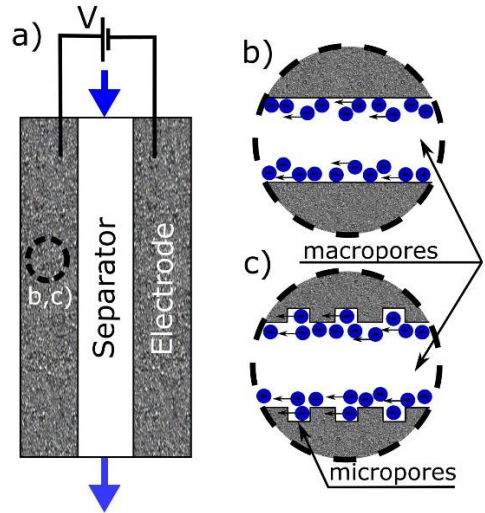

**Figure 1: a) Schematic of a typical CDI cell with feedwater flow between the two electrodes. Schematics b) and c) show the pore structures investigated in this work, which include b) solely through-electrode macropores, and c) through-electrode macropores which access micropores.**

## Theory

To develop the theory of surface transport in CDI cells, we begin by describing transport of salt and charge in the electroneutral bulk of a macropore (see Figure 1). For a binary and symmetric electrolyte, non-dimensional salt and charge balance equations are given by:

$$\frac{\partial \hat{c}}{\partial \hat{t}} = \hat{\nabla}^2 \hat{c} \tag{1}$$

$$\hat{\nabla}\left(\hat{c}\hat{\nabla}\hat{\phi}\right) = 0 \tag{2}$$

Where $\hat{c}$ is local salt concentration in the pore bulk scaled by the initial pore bulk salt concentration, $c_o$, and $\hat{\phi}$ is the local bulk electrostatic potential non-dimensionalized by the thermal voltage. Further, $\hat{t}$ is time scaled by the characteristic diffusion time across the characteristic pore size, $h_p^2/D$, where $D$ is the diffusivity of both anion and cation and $h_p$ is the pore volume divided by pore surface area[37]. The parameter $h_p$ is used as the characteristic lengthscale for the scaled bulk coordinates $\hat{x}$, $\hat{y}$, and $\hat{z}$. During cell charging, counterions electromigrate towards macropore walls (Figure 1 b and c) or into micropores (Figure 1c) while coions electromigrate away. This results in the formation of non-electroneutral electric double layers (EDLs) along the macropore wall with a characteristic thickness given by the Debye length, $\lambda_D$, and also within micropores where the characteristic size is often on the order of $\lambda_D$. Including the EDLs in the model domain requires a solution of the coupled Poisson-Nernst-Plank equations, which can be numerically challenging as the EDLs and micropores are often significantly smaller than the macropore size, $h_p$. An alternative method for thin EDLs is to instead model the effects of EDL charging on the bulk domain via effective flux boundary conditions[32,33,38]. For the case of ion transport between a diffuse EDL and bulk domain, and surface (tangential) transport of ions within the diffuse EDL, Chu and Bazant presented the following general effective flux boundary conditions at the location of the EDL[32] (e.g. at $\hat{z}=0$):

$$\varepsilon \frac{\partial \tilde{w}}{\partial \hat{t}} = \varepsilon \hat{\nabla}_s \cdot \left( \hat{\nabla}_s \tilde{w} + \tilde{q} \hat{\nabla}_s \hat{\phi} + \int_0^\infty \tilde{\rho} \hat{\nabla}_s \tilde{\psi} d\tilde{z} \right) - \left( \frac{\partial \hat{c}}{\partial \hat{z}} \right) \bigg|_{(\hat{x},\hat{y},0)} \quad (3)$$

$$\varepsilon \frac{\partial \tilde{q}}{\partial \hat{t}} = \varepsilon \hat{\nabla}_s \cdot \left( \hat{\nabla}_s \tilde{q} + \tilde{w} \hat{\nabla}_s \hat{\phi} + \int_0^\infty \tilde{c} \hat{\nabla}_s \tilde{\psi} d\tilde{z} \right) - \left( \hat{c} \frac{\partial \hat{\phi}}{\partial \hat{z}} \right) \bigg|_{(\hat{x},\hat{y},0)} \quad (4)$$

Where the tilde refers to variables of the EDL (inner) domain. The EDL has inner coordinates $\tilde{x}$ and $\tilde{y}$ parallel to the pore surface and non-dimensionalized by $h_p$, and $\tilde{z}$ perpendicular to the surface and scaled by $\lambda_D$, with $\tilde{z}=0$ being the location of the solid wall. Further, $\hat{\nabla}_s$ is the surface gradient[38], $\varepsilon$ is the ratio $\lambda_D/h_p$, and $\tilde{\rho}$ is the local net charge density in the EDL, defined as $(\tilde{c}_c - \tilde{c}_a)/2$, where $\tilde{c}_a$ and $\tilde{c}_c$ are, respectively, the anion and cation concentration in the EDL non-dimensionalized by $c_o$. $\tilde{c}$ is the local mean ion concentration in the EDL, defined as $(\tilde{c}_c + \tilde{c}_a)/2$, and $\tilde{\psi}$ is the excess potential in the EDL[32]. Also, $\tilde{q}$ is the non-dimensional charge density in the EDL and $\tilde{w}$ is the non-dimensional EDL excess salt density in the EDL, defined as[33]:

$$\tilde{w} = \int_0^\infty (\tilde{c} - \hat{c}) d\tilde{z} \quad (5)$$

$$\tilde{q} = \int_0^\infty \tilde{\rho} d\tilde{z} \tag{6}$$

Additionally, Chu and Bazant derived effective flux boundary conditions for the case of a Gouy-Chapman (GC) EDL, which were given as the following[32,33]:

$$\tilde{w} = 4\sqrt{\hat{c}} \sinh^2 \frac{\tilde{\zeta}}{4} \tag{7}$$

$$\tilde{q} = -2\sqrt{\hat{c}} \sinh \frac{\tilde{\zeta}}{2} \tag{8}$$

$$\varepsilon \frac{\partial \tilde{w}}{\partial \hat{t}} = \varepsilon \hat{\nabla}_s \cdot \left( \tilde{w} \hat{\nabla}_s \ln \hat{c} + \tilde{q} \hat{\nabla}_s \hat{\phi} \right) - \left( \frac{\partial \hat{c}}{\partial \hat{z}} \right)\bigg|_{(\hat{x},\hat{y},0)} \tag{9}$$

$$\varepsilon \frac{\partial \tilde{q}}{\partial \hat{t}} = \varepsilon \hat{\nabla}_s \cdot \left( \tilde{q} \hat{\nabla}_s \ln \hat{c} + \tilde{w} \hat{\nabla}_s \hat{\phi} \right) - \left( \hat{c} \frac{\partial \hat{\phi}}{\partial \hat{z}} \right)\bigg|_{(\hat{x},\hat{y},0)} \tag{10}$$

Where $\tilde{\zeta}$ is the zeta potential, defined as the induced potential drop across the GC EDL. While a GC EDL can be used to approximate the EDL structure along macropore walls, such a model cannot describe EDL structure in micropores due to a characteristic size which is on the order of $\lambda_D$ [39]. Instead, for micropores a Donnan EDL model is used, which assumes uniform micropore electric potential and concentration, reflecting the geometric confinement and resulting strong EDL overlap in micropores[40–42]. Modified versions of the Donnan model have been used to fit model results to experimental data, where modifications include a Stern layer and an additional adsorption potential[43], or capture the amphoteric nature of the carbon electrode[44]. To model surface transport in CDI systems with micropores, we here derived the effective flux boundary conditions for the case of a Donnan EDL (see Appendix A for detailed derivation):

$$\varepsilon_{mi} \frac{\partial \left( \tilde{c}_{mi} - \hat{c} \right)}{\partial \hat{t}} = \varepsilon_{mi} \hat{\nabla}_s^2 \left( \tilde{c}_{mi} - \hat{c} \right) - \left( \frac{\partial \hat{c}}{\partial \hat{z}} \right)\bigg|_{(\hat{x},\hat{y},0)} \tag{11}$$

$$\varepsilon_{mi} \frac{\partial \tilde{\sigma}_{mi}}{\partial \hat{t}} = \varepsilon_{mi} \hat{\nabla}_s \cdot \left[ \hat{\nabla}_s \tilde{\sigma}_{mi} - \hat{c} \hat{\nabla}_s \hat{\phi} \right] - \left( \hat{c} \frac{\partial \hat{\phi}}{\partial \hat{z}} \right)\bigg|_{(\hat{x},\hat{y},0)} \tag{12}$$

Where $\varepsilon_{mi} = l_{mi}/h_p$, $l_{mi}$ is the depth of the micropores, $\tilde{c}_{mi} \equiv \left( \tilde{c}_{mi,c} + \tilde{c}_{mi,a} \right)/2$ is the micropore salt concentration, and $\tilde{\sigma}_{mi} \equiv \left( \tilde{c}_{mi,c} - \tilde{c}_{mi,a} \right)/2$ is the micropore charge density. We note that, in Eqs. (11) and (12), all of the micropores' excess salt and charge participates in surface transport, and thus these equations give an upper limit of surface transport due to micropore EDLs. A more refined model capturing precisely micropore surface transport requires resolving micropore geometric features. However, such a model is beyond the scope of the current work. We note that some CDI electrode materials with long micropores (relative to micropore hydraulic diameter), such as activated carbon electrodes consisting of microporous micron-scale carbon particles, may result in most of the micropore ionic charge being unavailable for surface transport. However others, such as hierarchical activated carbon aerogels (HCAMs), where short micropores are etched into macropore walls, may exhibit a more significant fraction of micropore charge and salt contributing to effective tangential fluxes.

Eqs. (1)-(2) with either Eqs. (9)-(10) or Eqs. (11)-(12) as boundary conditions can be applied to study the macropore salt and charge dynamics. However, these models require accounting for the complexity of the macropore geometry, which is often difficult for random porous media. To avoid geometric complexities, a common technique is to develop macroscopic models based on volume averaged variables[45]. As described in Biesheuvel and Bazant[37], via integrating the pore bulk transport equations, here Eqs. (1) and (2), over a suitable representative volume, the normal flux component of the effective flux boundary conditions can be converted to volumetric source terms in a macroscopic model. The latter authors developed such volumetric source terms for the case of ion electrosorption into GC EDLs but did not include the effect of surface transport within the EDL[37]. We here extend the latter work to include surface transport via volume integrating the pore bulk transport equations and implementing the effective flux boundary conditions given by Eqs. (9) and (10). The complete procedure is shown in Appendix B, and we here present the results for the one-dimensional form of the macroscopic transport equations:

$$\frac{\partial(\bar{c}+\varepsilon\bar{w})}{\partial \bar{t}} = \frac{\partial^2 \bar{c}}{\partial \bar{x}^2} + \varepsilon \frac{\partial}{\partial \bar{x}}\left(\bar{w}\frac{\partial \ln \bar{c}}{\partial \bar{x}} + \bar{q}\frac{\partial \bar{\phi}}{\partial \bar{x}}\right) \quad (13)$$

$$\varepsilon \frac{\partial \bar{q}}{\partial \bar{t}} = \frac{\partial}{\partial \bar{x}}\left(\bar{c}\frac{\partial \bar{\phi}}{\partial \bar{x}}\right) + \varepsilon \frac{\partial}{\partial \bar{x}}\left(\bar{q}\frac{\partial \ln \bar{c}}{\partial \bar{x}} + \bar{w}\frac{\partial \bar{\phi}}{\partial \bar{x}}\right) \quad (14)$$

We denote Eq. (13) and (14) as our 1D surface transport (ST) model, and $\bar{c}$ and $\bar{\phi}$ represent volume-averaged bulk concentration and potential, respectively, while $\bar{w}$ and $\bar{q}$ are the surface area averaged excess salt concentration and charge, respectively. The variables $\bar{x}$ and $\bar{t}$ are defined as $\bar{x} = x/L_e$ and $\bar{t} = tD/L_e^2$, where $L_e$ is the electrode's thickness. As shown in Equations (13) and (14), the tangential flux components of the effective flux boundary conditions have been converted to volumetric flux terms in the macroscopic equations. We can compare our 1D ST model to the 1D model presented by Mirzadeh et al. for surface transport in a supercapacitor electrode (Eq. 12 and 13 of that work),[35] which we reproduce here:

$$\frac{\partial(\bar{c}+\varepsilon\bar{w})}{\partial \bar{t}} = \frac{\partial^2(\bar{c}+\varepsilon\bar{w})}{\partial \bar{x}^2} + \varepsilon \frac{\partial}{\partial \bar{x}}\left(\bar{q}\frac{\partial \bar{\phi}}{\partial \bar{x}}\right) \quad (15)$$

$$\varepsilon \frac{\partial \overline{q}}{\partial \overline{t}} = \varepsilon \frac{\partial^2 \overline{q}}{\partial \overline{x}^2} + \frac{\partial}{\partial \overline{x}}\left((\overline{c} + \varepsilon \overline{w})\frac{\partial \overline{\phi}}{\partial \overline{x}}\right) \qquad (16)$$

As can be seen, our 1D ST model differs from the Mirzadeh model, as the latter model did not include the effect of surface conduction driven by gradients in the excess potential. In other words, Eqs. (15) and (16) can be derived by volume integrating the bulk equations and implementing Eqs. (3) and (4) with $\hat{\nabla}_s \hat{\psi} = 0$ as effective flux boundary conditions, as we show in Appendix B. All models in this work were solved using COMSOL Multiphysics 4.4 (COMSOL Inc., Sweden).

## Results

In Figure 2a, we compare the charging dynamics predicted by the Mirzadeh model (Eqs. (15) and (16)) to our 1D ST model (Eqs. (13) and (14)), and to the 1D model without surface transport presented by Biesheuvel and Bazant (the BB model)[37]. In this figure, $q$ represents the time-dependent charge stored in the electrode after applying the voltage, and $q_{ss}$ is the charge stored at steady state. For the time axis, we non-dimensionalized time by the transmission line timescale $\tau_{TL} = \lambda_D L_e^2 / h_p D$ [35,46]. To contrast our 1D ST model to the Mirzadeh model, we investigate the simpler case where desalination is inhibited during cell charging by using boundary conditions of $\overline{c} = 1$ and $\overline{\phi} = 0$ at the inner electrode edge, $\overline{x} = 0$. We note that this concentration boundary condition, which was also used in the work of Mirzadeh et al.,[35,47] is most appropriate for modeling supercapacitor systems where salt concentration in the electrode is not expected to vary during cell charging. We further used a zero flux boundary condition (both bulk and surface flux) at the outer electrode edge, $\overline{x} = 1$, and initial conditions of $\overline{c} = 1$ and $\overline{\phi} = V_{el} = 7.5$ along the length of the electrode, where $V_{el}$ is the voltage applied to the solid phase of the electrode. As seen in Figure 2a, including surface transport leads to significantly faster dynamics in both the Mirzadeh model (as previously reported[35]) and our 1D ST model, when compared to the BB model. However, we can also observe that our 1D ST model demonstrates slower charging compared to the Mirzadeh model for most of the charging process. The latter is due to the inclusion in the 1D ST model of surface transport driven by gradients in the excess potential. The excess electric field is in the opposite direction of the bulk electric field, as is shown in Figure 2b where we plot excess potential at $\tilde{z} = 0$, which we denote as $\tilde{\psi}_w$ and note that $\tilde{\psi}_w = \tilde{\zeta}$ (blue lines), and also plot the bulk potential, $\overline{\phi}$ (black lines).

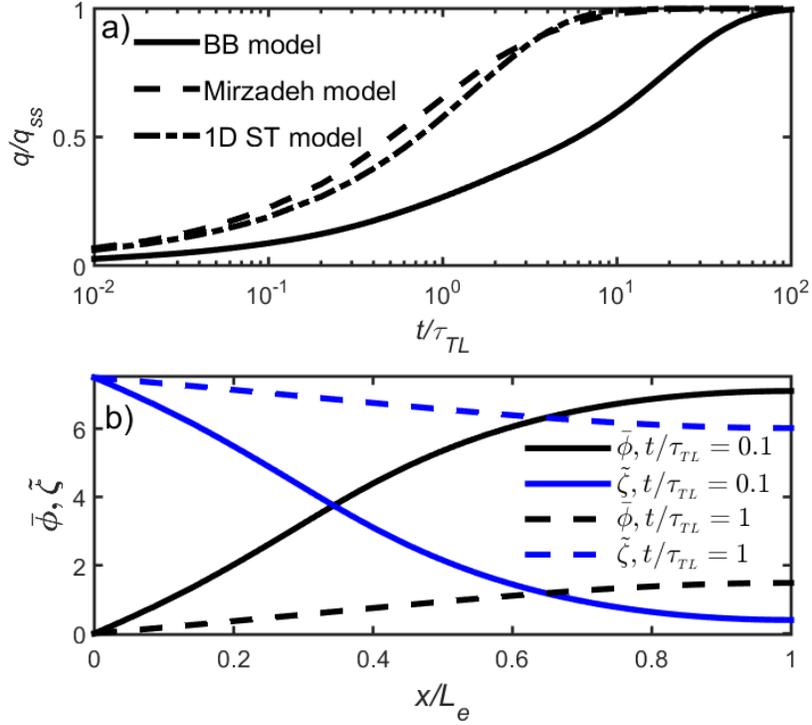

**Figure 2: a)** Predicted electrode charge stored, $q/q_{ss}$, of our 1D ST model, the Mirzadeh model[35], and the BB model[37] with supercapacitor boundary conditions and $\varepsilon = 0.1$, $V_{el} = 7.5$. **b)** Plot of the bulk potential, $\hat{\phi}$, and excess potential at the pore wall, $\tilde{\psi}_w = \tilde{\zeta}$, predicted by the 1D ST model.

We now utilize our 1D ST model to develop insight into the effect of surface transport on the cell charging and salt dynamics in a CDI cell, thus restricting ourselves for the moment to solely macroporous CDI electrodes with GC EDLs along the pore walls (Figure 1b). For predicting CDI cell performance, appropriate boundary conditions must be used which allow for desalination at the electrode's inner edge. Thus, we solve our 1D ST model for cells which include a porous separator layer adjacent to the porous electrode, and for two physical scenarios in the separator, see Figure 3a and b. For simplicity the separator layer's porosity is set to be equal to that of the electrode. The first scenario, shown schematically in Figure 3a, models the case of a stagnant diffusion layer (SDL) adjacent to the electrode, which has been used previously to model CDI cells[37]. The governing equations in the SDL are:

$$\frac{\partial \bar{c}}{\partial \bar{t}} = \frac{\partial^2 \bar{c}}{\partial \bar{x}^2} \tag{17}$$

$$\frac{\partial}{\partial \bar{x}}\left(\bar{c}\frac{\partial \bar{\phi}}{\partial \bar{x}}\right) = 0 \tag{18}$$

and the boundary conditions at the edge of the separator are:

$$\bar{c}\big|_{\bar{x}=-l_{SDL}/L_e=-0.05} = 1, \bar{\phi}\big|_{\bar{x}=-0.05} = 0 \tag{19}$$

The second case, shown schematically in Figure 3b, is the case of a CDI cell which is symmetric about its midline (the midline of the separator). Such a case can approximate the physical scenario of batch mode operation whereby a batch of feedwater in the cell is desalted while flow is turned off[9]. For the symmetric cell, the boundary conditions at the midline of the separator are:

$$\left.\frac{\partial \bar{c}}{\partial \bar{x}}\right|_{\bar{x}=-0.05} = 0, \left.\bar{\phi}\right|_{\bar{x}=-0.05} = 0 \qquad (20)$$

For both cases (SDL and symmetric cell), at the electrode's outer edge we applied zero flux boundary conditions. The initial conditions for both cases are $\bar{c}=1$ throughout the electrode and separator, $\bar{\phi}=7.5$ in the electrode, and a linear potential profile in the separator from $\bar{\phi}=0$ at $\bar{x}=-0.05$ to $\bar{\phi}=7.5$ at $\bar{x}=0$.

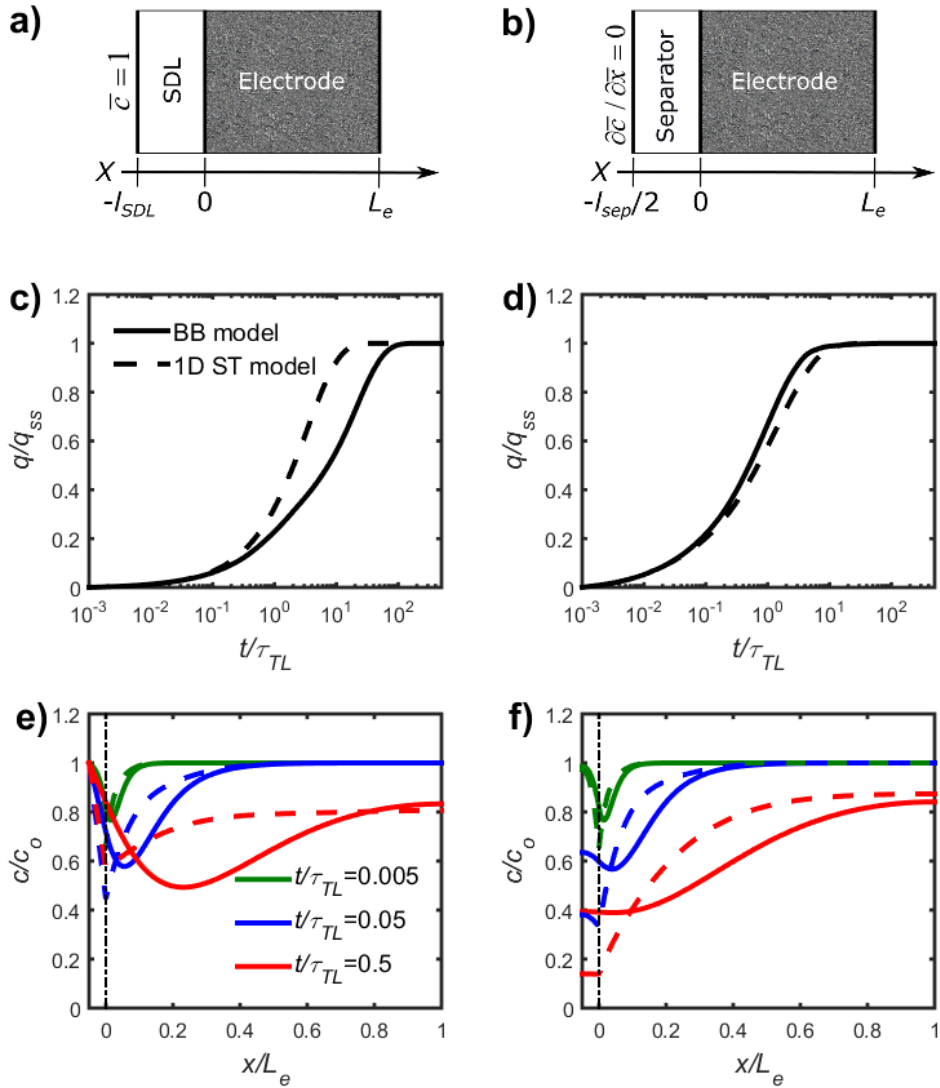

Figure 3: a) A schematic of the electrode with adjacent stagnant diffusion layer (SDL). c) And e) show results from the cell with an adjacent SDL layer, using either our 1D ST model or the BB model to describe the electrode. c)

Shows non-dimensional electrode charge stored, $q/q_{ss}$, versus time scaled by the transmission line timescale, $\tau_{TL}$, and e) shows the concentration in the electrode and SDL, $\bar{c}$, versus non-dimensional position, $x/L_e$. b) A schematic of the symmetric cell, and d) and f) show results using either our 1D ST model or the BB model to describe the electrode. All model results are for $\varepsilon = 0.1$, $V_{el} = 7.5$, $l_{SDL} = l_{sep}/2 = 0.05 L_e$.

For the case of a CDI electrode with an adjacent SDL, we can see from Figure 3c that the 1D ST model's predicted charging (dashed line) is significantly faster than the BB model which does not consider surface transport (solid line), as might be expected given the results of Figure 2a. Figure 3e shows the concentration profiles and desalination dynamics at various times during cell charging for the 1D ST model (dashed lines) and the BB model (solid lines) for the cell with adjacent SDL. Here, we can observe a sharp discontinuity in the slope of the concentration profile at the separator/electrode interface (vertical dashed line) for the 1D ST model. The latter is due to the discontinuity in transport mechanisms across this interface, with surface transport occurring on the electrode side but not on the SDL side. For both the 1D ST and BB models, we see the expected result that EDL charging and so desalination begins at the electrode/SDL interface and then at later times propagates deeper into the electrode.[37] Conversely, for the symmetric cell, the 1D ST model predicts slower cell charging compared to the BB model, see Figure 3d. The latter results are highly counter-intuitive, as they suggest that the inclusion of an extra transport pathway for ions (surface transport) results in slower cell charging. These counter-intuitive results can be understood through the concentration profiles for the symmetric cell shown in Figure 3f. Here, we can see that during charging, as no ions can enter the separator space from the boundary at $\bar{x} = -0.05$ due to symmetry, the separator space ($\bar{x} < 0$) is strongly desalted relative to results of Figure 3e. The desalination of the separator space is especially strong for the 1D ST model in Figure 3f, where concentration in the separator space can be reduced by approximately an order of magnitude (see the red dashed curve). This region of very low ion concentration acts to slow down the charging dynamics by reducing the local ionic current in the separator space, as seen by Eq. (18), or in other words by introducing a large resistor to ionic current into the system. In summary, the presence of surface transport in the symmetric cell leads to a more effective desalting of the separator space, which in turn significantly slows the electrode charging.

As the 1D ST model employs a GC EDL, it simulates an electrode consisting solely of macropores with thin EDLs, as depicted schematically in Figure 1b. However, CDI electrodes are typically multiscale, consisting of both through-electrode macropores which access micropores with strongly overlapped EDLs, see Figure 1c. Thus, to capture the multiscale nature of CDI electrodes, we developed a 2D model of a single slit-shaped macropore with uniform cross-sectional area, and with micropores present along the macropore wall at uniform intervals, as drawn schematically in Figure 1c. For this model, we used Eqs. (1) and (2) to govern the macropore bulk dynamics, and used an alternating arrangement of the GC effective flux boundary conditions for macropore walls, and the Donnan effective flux boundary conditions for micropores, to describe electrosorption into EDLs and surface transport within EDLs. We do not develop a 1D macroscopic set of equations for the case of multiscale electrodes, as such an approach would average over the serial nature of the EDL arrangement. As we show below, this local serial arrangement can have important implications in the cell charging and desalination dynamics. The importance of capturing local changes in surface conditions in charging porous electrodes was also shown by Mirzadeh et al., for the case of a "patchy" electrode with alternating GC EDLs and zero surface charge areas.[35]

For the 2D model, the macropore length used was $L_e = 100 h_p$, and the alternating arrangement of Donnan (micropore) and GC (macropore wall) effective flux boundary conditions was located along the $\hat{y} = 1$ boundary. The first micropore was positioned at $\hat{x} = 0.2$, the width of all micropores was $\Delta \hat{x} = 0.1$, and the distance between neighboring micropores was $\Delta \hat{x} = 0.4$. We added an SDL layer adjacent to the pore inlet with thickness $L_e / 20$, used a symmetry condition along the macropore midline ($\hat{y} = 0$), and imposed no flux boundary conditions at the pore closed end. The initial conditions used were $\hat{c} = 1$ and $\hat{\phi} = V_{el} = 12.5$ in the macropore, and a linear potential profile in the SDL along the $\hat{x}$ axis from $\hat{\phi} = 0$ at $\hat{x} = -0.05$ to $\hat{\phi} = 12.5$ at $\hat{x} = 0$. In Figure 4a, we show the charging dynamics of three 2D models versus time non-dimensionalized by the diffusion time $\tau_D = L_e^2 / D$. The three models include the "macropore & micropore ST" model where the macropore wall and micropore EDLs were modeled using the boundary conditions (9)-(12), the "macropore ST" model where the terms including the surface gradient, $\hat{\nabla}_s$. were dropped out of Eqs. (11)-(12), and the "no ST" model where the terms including $\hat{\nabla}_s$ were dropped out of Eqs. (9)-(12). As can be seen, the "macropore & micropore ST" model predicts the fastest charging, which is consistent with the results of Figure 3c where it was demonstrated that surface transport enhances CDI cell charging kinetics when including an adjacent SDL. Counter-intuitively, the "macropore ST" model shows slower cell charging than the "no ST" model. Here, the salt dynamics can shed insight into the observed counter-intuitive behavior. To this end, Figure 4b-j shows the 2D concentration fields in the macropore inlet ($0 \leq \hat{x} \leq 2$) for each model at various charging times. As can be seen in Figure 4b-d, at the time $t / \tau_D = 1E - 7$ desalination zones appear associated with the location of the micropores in all three model cases, with the strongest desalination occurring closest to the pore entrance. Figure 4e-g demonstrate that at a later time $t / \tau_D = 1E - 4$, significant differences between the three models can be observed, where a large depletion region is seen near to $\hat{y} = 1$ for the "macropore ST" model relative to that of the other two model cases. This latter depletion region can act to slow down the macropore charging, due to the introduction of local areas of high ionic resistance within the macropore, explaining the counter-intuitive results of Figure 4a. In Figure 4h-j, we see that at the even later time $t / \tau_D = 1E - 3$, we can see that the pore entrance is now more uniformly desalted, with the lowest concentration again observed for the "macropore ST" model.

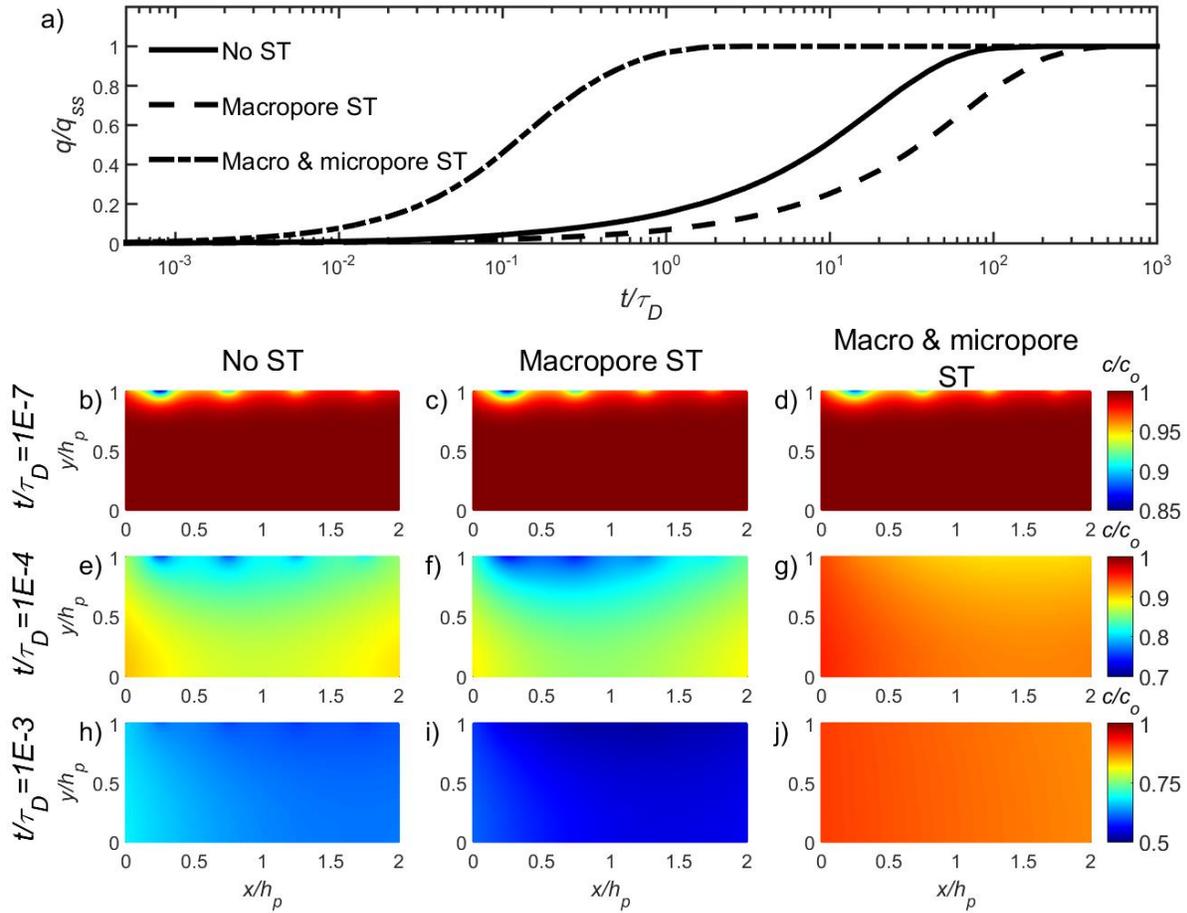

**Figure 4: a)** Model results for non-dimensional charge stored in EDLs, $q/q_{ss}$, **versus time scaled by the diffusion timescale,** $\tau_D$, **for three 2D models, including the "No ST" model (solid line), the "macropore ST" model (dashed line), and the "macro & micropore ST" model (dotted-dashed line). b-d) show model results of concentration distributions in the macropore entrance at** $t/\tau_D = 1E-7$, **e-g) show the concentration distributions for** $t/\tau_D = 1E-4$, **and h-j) show the concentration distributions** $t/\tau_{Diff} = 1E-3$. **The parameters used for these results include** $\varepsilon = 0.001$, $\varepsilon_{mi} = 0.001$, **and** $V_{el} = 12.5$.

## Conclusions

In conclusion, we developed theory which includes surface transport in describing the charge and salt dynamics in a CDI cell. We presented an effective flux boundary condition for micropore Donnan EDLs, a set of 1D macroscopic transport equations for macroporous CDI electrodes, and a 2D model for a multi-scale CDI electrode. Through these models, the possible effect of surface transport on the CDI process is elucidated. It was observed that at some model conditions surface transport enhances cell charging rates, but counter-intuitively this additional transport pathway is found to slow down the cell charging at other model conditions. While we here begin to explore the effects of surface transport in CDI cells, future models should include additional complexities necessary to accurately predict data from CDI cells, such as the effect of the Stern layer, and geometric as well as steric effects in the highly constrained micropores.

**Acknowledgements:**
We would like to acknowledge Mathias Baekbo Anderson for insightful discussions, and funding from the Israel National Research Center for Electrochemical Propulsion (INREP).

**Appendix A:**
In this appendix, we derive effective flux boundary conditions for the case of micropores described by a Donnan EDL. The Donnan model assumes a spatially uniform potential within the micropore. Via a Boltzmann distribution, the Donnan potential drop between the micropore and macropore can be related to the micropore concentration[42]:

$$\tilde{c}_{mi,i} = \hat{c} e^{-z_i \Delta \hat{\phi}_D} \tag{A1}$$

Here $\tilde{c}_{mi,i}$ is the concentration of ion $i$ in the micropore non-dimensionalized by $c_o$, $z_i$ is the valance of ion $i$, and $\Delta \hat{\phi}_D$ is the Donnan potential non-dimensionalized by the thermal voltage. Next, we introduce the parameters $\tilde{c}_{mi}$ and $\tilde{\sigma}_{mi}$ representing, respectively, micropore salt concentration and charge density:

$$\tilde{c}_{mi} \equiv \frac{\tilde{c}_{mi,c} + \tilde{c}_{mi,a}}{2} = \hat{c} \cosh \Delta \hat{\phi}_D \tag{A2}$$

$$\tilde{\sigma}_{mi} \equiv \frac{\tilde{c}_{mi,c} - \tilde{c}_{mi,a}}{2} = -\hat{c} \sinh \Delta \hat{\phi}_D \tag{A3}$$

Where the expressions developed in (A2) and (A3) are valid for the case of monovalent ions. We also notice that in the micropores:

$$\tilde{\psi} = \Delta \hat{\phi}_D \tag{A4}$$

In order to substitute micropore quantities into the general effective flux boundary conditions given by Chu and Bazant,[32] and here as Eqs. (3) and (4), we notice that:

$$\varepsilon \tilde{w} = \frac{\lambda_D}{h_p} \int_0^\infty (\tilde{c} - \hat{c}) d\tilde{z} = \frac{1}{h_p} \int_0^{l_{mi}} (\tilde{c} - \hat{c}) dz = \varepsilon_{mi} (\tilde{c}_{mi} - \hat{c}) \tag{A5}$$

$$\varepsilon \tilde{q} = \frac{\lambda_D}{h_p} \int_0^\infty \tilde{\rho} d\tilde{z} = \frac{1}{h_p} \int_0^{l_{mi}} \tilde{\rho} dz = \varepsilon_{mi} \tilde{\sigma}_{mi} \tag{A6}$$

$$\varepsilon \int_0^\infty \tilde{\rho} \hat{\nabla}_s \tilde{\psi} d\tilde{z} = \varepsilon_{mi} \tilde{\sigma}_{mi} \hat{\nabla}_s \left( \Delta \hat{\phi}_D \right) \tag{A7}$$

$$\varepsilon \int_0^\infty \tilde{c} \hat{\nabla}_s \tilde{\psi} d\tilde{z} = \varepsilon_{mi} \tilde{c}_{mi} \hat{\nabla}_s \left( \Delta \hat{\phi}_D \right) \tag{A8}$$

Substituting (A5)-(A8) into (3) and (4) lead to:

$$\varepsilon_{mi} \frac{\partial (\tilde{c}_{mi} - \hat{c})}{\partial \hat{t}} = \varepsilon_{mi} \hat{\nabla}_s \left[ \hat{\nabla}_s (\tilde{c}_{mi} - \hat{c}) + \tilde{\sigma}_{mi} \hat{\nabla}_s \left( \hat{\phi} + \Delta \hat{\phi}_D \right) \right] - \left( \frac{\partial \hat{c}}{\partial \hat{z}} \right)\bigg|_{(\hat{x},\hat{y},0)} \tag{A9}$$

$$\varepsilon_{mi} \frac{\partial \tilde{\sigma}_{mi}}{\partial \hat{t}} = \varepsilon_{mi} \hat{\nabla}_s \left[ \hat{\nabla}_s \tilde{\sigma}_{mi} + \tilde{c}_{mi} \hat{\nabla}_s \left( \hat{\phi} + \Delta \hat{\phi}_D \right) - \hat{c} \hat{\nabla}_s \hat{\phi} \right] - \left( \hat{c} \frac{\partial \hat{\phi}}{\partial \hat{z}} \right)\bigg|_{(\hat{x},\hat{y},0)} \tag{A10}$$

For the case where we neglect the Stern layer, we can simplify by noting that the Donnan potential drop equals the potential drop between the electrode surface and the macropore, such that:

$$\hat{\nabla}_s \left( \hat{\phi} + \Delta \hat{\phi}_D \right) = \hat{\nabla}_s \left( \hat{\phi} + \hat{\phi}_{el} - \hat{\phi} \right) = \hat{\nabla}_s \left( \hat{\phi}_{el} \right) = 0 \tag{A11}$$

Where $\hat{\phi}_{el}$ is the non-dimensional solid phase potential of the electrode, which we assume is held constant throughout the charging process. Substituting (A11) into (A9) and (A10) lead to:

$$\varepsilon_{mi} \frac{\partial (\tilde{c}_{mi} - \hat{c})}{\partial \hat{t}} = \varepsilon_{mi} \hat{\nabla}_s^2 (\tilde{c}_{mi} - \hat{c}) - \left( \frac{\partial \hat{c}}{\partial \hat{z}} \right)\bigg|_{(\hat{x},\hat{y},0)} \tag{A12}$$

$$\varepsilon_{mi} \frac{\partial \tilde{\sigma}_{mi}}{\partial \hat{t}} = \varepsilon_{mi} \hat{\nabla}_s \left[ \hat{\nabla}_s \tilde{\sigma}_{mi} - \hat{c} \hat{\nabla}_s \hat{\phi} \right] - \left( \hat{c} \frac{\partial \hat{\phi}}{\partial \hat{z}} \right)\bigg|_{(\hat{x},\hat{y},0)} \tag{A13}$$

# Appendix B

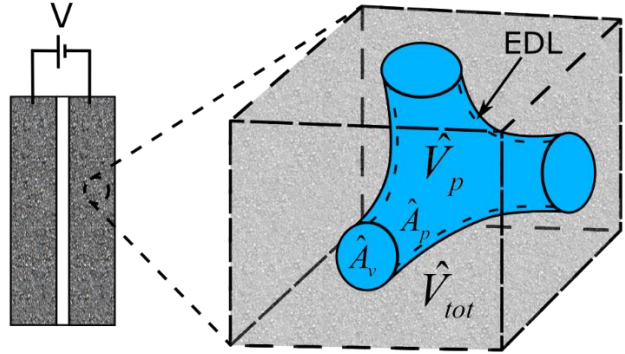

**Figure 5: Schematic of a representative volume element used in the volume averaging process, with non-dimensional parameters shown. $\hat{A}_v$ is the non-dimensional cross-sectional area of the pores, $\hat{V}_p$ is the pore volume, $\hat{A}_p$ is the pore surface area, and $\hat{V}_{tot}$ is the total volume of the element including the solid phase. Adapted from Figure 6 in Biesheuvel et al.[37]**

We here derive the macroscopic transport equations of the 1D ST model, Eqs. (13) and (14). We begin by the volume integration of the local transport equations, Eqs. (1) and (2), over the non-dimensional pore volume, $\hat{V}_p$, of the characteristic volume element presented in Figure 5. The volume element consists of a solid phase and a liquid phase with a bulk electrolyte and thin EDL. We begin by integrating Eqs. (1) and (2) over $\hat{V}_p$, and applying the Divergence theorem[37,45]:

$$\int_{\hat{V}_p} \frac{\partial \hat{c}}{\partial \hat{t}} d\hat{V} = \int_{\hat{V}_p} \hat{\nabla}^2 \hat{c} d\hat{V} = \left[ \int_{\hat{A}_v} \hat{n} \cdot \left( \hat{\nabla} \hat{c} \right) d\hat{A} + \int_{\hat{A}_p} \hat{n} \cdot \left( \hat{\nabla} \hat{c} \right) d\hat{A} \right] \quad \text{(B1)}$$

$$0 = \int_{\hat{V}_p} \hat{\nabla} \left( \hat{c} \hat{\nabla} \hat{\phi} \right) d\hat{V} = \left[ \int_{\hat{A}_v} \hat{n} \cdot \left( \hat{c} \hat{\nabla} \hat{\phi} \right) d\hat{A} + \int_{\hat{A}_p} \hat{n} \cdot \left( \hat{c} \hat{\nabla} \hat{\phi} \right) d\hat{A} \right] \quad \text{(B2)}$$

Where $\hat{n}$ is the inwards direction normal to a surface, the dimensionless areas and volumes are defined by $\hat{A} = A/h_p^2$, and $\hat{V} = V/h_p^3$, and $\hat{A}_v$ represents the cross-sectional area of the pores and $\hat{A}_p$ the pore surface area.

Next we define volume-averaged bulk concentration $\bar{c}$ and potential, $\bar{\phi}$, as:

$$\bar{c} = \frac{1}{\hat{V}_p} \int_{\hat{V}_p} \hat{c} d\hat{V} \quad \text{(B3)}$$

$$\bar{\phi} = \frac{1}{\hat{V}_p} \int_{\hat{V}_p} \hat{\phi} d\hat{V} \tag{B4}$$

We also define area-averaged EDL quantities, such as surface area-averaged excess salt concentration, $\bar{w}$, and charge, $\bar{q}$ [37]:

$$\bar{w} = \frac{1}{\hat{A}_p} \int_{\hat{A}_p} \tilde{w} d\hat{A} \tag{B5}$$

$$\bar{q} = \frac{1}{\hat{A}_p} \int_{\hat{A}_p} \tilde{q} d\hat{A} \tag{B6}$$

We can then rewrite the first term of Eq. (B1) using (B3) and re-scaling time to the electrode diffusion time, $\bar{t} = tD/L_e^2$:

$$\int_{\hat{V}_p} \frac{\partial \hat{c}}{\partial \hat{t}} d\hat{V} = \hat{V}_p \frac{\partial \bar{c}}{\partial \hat{t}} = \frac{h_p^2 \hat{V}_p}{L_e^2} \cdot \frac{\partial \bar{c}}{\partial \bar{t}} \tag{B7}$$

The cross-sectional flux terms of Eqs. (B1) and (B2) can be rewritten using volume averaged parameters in the following way:

$$\int_{\hat{A}_v} \hat{n} \cdot (\hat{\nabla}\hat{c}) d\hat{A} = \int_{\hat{A}_v + \hat{A}_s} \hat{n} \cdot (\hat{\nabla}\hat{c}) d\hat{A} = \int_{\hat{V}_{tot}} \hat{\nabla}^2 \hat{c} d\hat{V} = \frac{h_p^2 \hat{V}_p}{L_e^2} \bar{\nabla}^2 \bar{c} \tag{B8}$$

$$\int_{\hat{A}_v} \hat{n} \cdot (\hat{c}\hat{\nabla}\hat{\phi}) d\hat{A} = \int_{\hat{A}_v + \hat{A}_s} \hat{n} \cdot (\hat{c}\hat{\nabla}\hat{\phi}) d\hat{A} = \int_{\hat{V}_{tot}} \hat{\nabla} \cdot (\hat{c}\hat{\nabla}\hat{\phi}) d\hat{V} \sim \frac{h_p^2 \hat{V}_p}{L_e^2} \bar{\nabla}(\bar{c}\bar{\nabla}\bar{\phi}) \tag{B9}$$

Where we rescaled the gradient, divergence and Laplacian operators with the electrode thickness, $L_e$. Further, in Eq. (B9) we assumed that variables are slowly varying at the pore-scale, and thus $\int_{\hat{V}_{tot}} \hat{c}\hat{\nabla}\hat{\phi} d\hat{V} \sim \hat{V}_p \bar{c}\bar{\nabla}\bar{\phi}$ [37].

Next we rewrite the last term in the right hand side of Eqs. (B1) and (B2) using the effective flux boundary conditions, Eqs. (9) and (10), and again assume slowly-varying variables to obtain:

$$\int_{\hat{A}_p} \hat{n} \cdot (\hat{\nabla}\hat{c}) d\hat{A} \sim \frac{\varepsilon \hat{A}_p h_p^2}{L_e^2} \left[ \bar{\nabla}_s \left( \bar{w}\bar{\nabla}_s \ln \bar{c} + \bar{q}\bar{\nabla}_s \bar{\phi} \right) - \frac{\partial \bar{w}}{\partial \bar{t}} \right] \tag{B10}$$

$$\int_{\hat{A}_p} \hat{n} \cdot (\hat{c}\hat{\nabla}\hat{\phi}) d\hat{A} \sim \frac{\varepsilon \hat{A}_p h_p^2}{L_e^2} \left[ \bar{\nabla}_s \left( \bar{q}\bar{\nabla}_s \ln \bar{c} + \bar{w}\bar{\nabla}_s \bar{\phi} \right) - \frac{\partial \bar{q}}{\partial \bar{t}} \right] \tag{B11}$$

We can now place Eqs. (B7)-(B11) into (B1) and (B2), and formulate our 1D macroscopic transport equations:

$$\hat{V}_p \frac{\partial \bar{c}}{\partial \bar{t}} = \hat{V}_p \frac{\partial^2 \bar{c}}{\partial \bar{x}^2} + \varepsilon \hat{A}_p \left[ \frac{\partial}{\partial \bar{x}} \left( \bar{w} \frac{\partial \ln \bar{c}}{\partial \bar{x}} + \bar{q} \frac{\partial \bar{\phi}}{\partial \bar{x}} \right) - \frac{\partial \bar{w}}{\partial \bar{t}} \right] \tag{B12}$$

$$0 = \hat{V}_p \frac{\partial}{\partial \bar{x}} \left( \bar{c} \frac{\partial \bar{\phi}}{\partial \bar{x}} \right) + \varepsilon \hat{A}_p \left[ \frac{\partial}{\partial \bar{x}} \left( \bar{q} \frac{\partial \ln \bar{c}}{\partial \bar{x}} + \bar{w} \frac{\partial \bar{\phi}}{\partial \bar{x}} \right) - \frac{\partial \bar{q}}{\partial \bar{t}} \right] \tag{B13}$$

Using the definition $h_p \equiv V_p / A_p$, we find that:

$$\frac{\hat{A}_p}{\hat{V}_p} = \frac{A_p}{h_p^2} \cdot \frac{h_p^3}{V_p} = 1 \tag{B14}$$

Thus, we can re-write (B12) and (B13) as:

$$\frac{\partial (\bar{c} + \varepsilon \bar{w})}{\partial \bar{t}} = \frac{\partial^2 \bar{c}}{\partial \bar{x}^2} + \varepsilon \frac{\partial}{\partial \bar{x}} \left( \bar{w} \frac{\partial \ln \bar{c}}{\partial \bar{x}} + \bar{q} \frac{\partial \bar{\phi}}{\partial \bar{x}} \right) \tag{B15}$$

$$\varepsilon \frac{\partial \bar{q}}{\partial \bar{t}} = \frac{\partial}{\partial \bar{x}} \left( \bar{c} \frac{\partial \bar{\phi}}{\partial \bar{x}} \right) + \varepsilon \frac{\partial}{\partial \bar{x}} \left( \bar{q} \frac{\partial \ln \bar{c}}{\partial \bar{x}} + \bar{w} \frac{\partial \bar{\phi}}{\partial \bar{x}} \right) \tag{B16}$$

Where Eqs. (B15) and (B16) are identical to Eqs. (13) and (14).

We here also briefly explain how to obtain the Mirzadeh model,[35] Eqs. (15) and (16), via volume averaging and utilizing effective flux boundary conditions. To begin, we re-write the effective flux boundary conditions, Eqs. (3) and (4), but neglecting the excess potential term:

$$\varepsilon \frac{\partial \tilde{w}}{\partial \hat{t}} = \varepsilon \hat{\nabla}_s \cdot \left( \hat{\nabla}_s \tilde{w} + \tilde{q} \hat{\nabla}_s \hat{\phi} \right) - \left. \frac{\partial \hat{c}}{\partial \hat{z}} \right|_{(\hat{x},\hat{y},0)} \tag{B17}$$

$$\varepsilon \frac{\partial \tilde{q}}{\partial \hat{t}} = \varepsilon \hat{\nabla}_s \cdot \left( \hat{\nabla}_s \tilde{q} + \tilde{w} \hat{\nabla}_s \hat{\phi} \right) - \left. \left( \hat{c} \frac{\partial \hat{\phi}}{\partial \hat{z}} \right) \right|_{(\hat{x},\hat{y},0)} \tag{B18}$$

Beginning with volume integrated bulk Eqs. (B1) and (B2), but instead using Eqs. (B17) and (B18) as the effective flux boundary conditions, we can derive the model presented by Mirzadeh et al.[35]:

$$\frac{\partial (\bar{c} + \varepsilon \bar{w})}{\partial \bar{t}} = \frac{\partial^2 \bar{c}}{\partial \bar{x}^2} + \varepsilon \frac{\partial}{\partial \bar{x}} \left( \frac{\partial \bar{w}}{\partial \bar{x}} + \bar{q} \frac{\partial \bar{\phi}}{\partial \bar{x}} \right) \tag{B19}$$

$$\varepsilon \frac{\partial \bar{q}}{\partial \bar{t}} = \frac{\partial}{\partial \bar{x}} \left( \bar{c} \frac{\partial \bar{\phi}}{\partial \bar{x}} \right) + \varepsilon \frac{\partial}{\partial \bar{x}} \left( \frac{\partial \bar{q}}{\partial \bar{x}} + \bar{w} \frac{\partial \bar{\phi}}{\partial \bar{x}} \right) \tag{B20}$$

# References

1. Suss, M. E. *et al.* Water desalination via capacitive deionization: what is it and what can we expect from it? *Energy Environ. Sci.* **8,** 2296–2319 (2015).

2. Porada, S., Feng, G., Suss, M. E. & Presser, V. Capacitive deionization in organic solutions : case study using propylene carbonate. *RSC Adv.* **6,** 5865–5870 (2016).

3. Seo, S. J. *et al.* Investigation on removal of hardness ions by capacitive deionization (CDI) for water softening applications. *Water Res.* **44,** 2267–2275 (2010).

4. Zhao, R. *et al.* Time-dependent ion selectivity in capacitive charging of porous electrodes. *J. Colloid Interface Sci.* **384,** 38–44 (2012).

5. Roelofs, S. H. *et al.* Capacitive deionization on-chip as a method for microfluidic sample preparation. *Lab Chip* **15,** 1458–1464 (2015).

6. Jeon, S., Yeo, J., Yang, S., Choi, J. & Kim, D. K. Ion storage and energy recovery of a flow-electrode capacitive deionization process. *J. Mater. Chem. A* **2,** 6378 (2014).

7. Zhao, R., Biesheuvel, P. M. & van der Wal, a. Energy consumption and constant current operation in membrane capacitive deionization. *Energy Environ. Sci.* **5,** 9520 (2012).

8. Biesheuvel, P. M., Hamelers, H. V. M. & Suss, M. E. Theory of Water Desalination by Porous Electrodes with Immobile Chemical Charge. *Colcom* **9,** 1–5 (2015).

9. Suss, M. E. *et al.* Capacitive desalination with flow-through electrodes. *Energy Environ. Sci.* **5,** 9511 (2012).

10. Cohen, I., Avraham, E., Bouhadana, Y., Soffer, A. & Aurbach, D. The effect of the flow-regime, reversal of polarization, and oxygen on the long term stability in capacitive de-ionization processes. *Electrochim. Acta* **153,** 106–114 (2015).

11. Avraham, E., Noked, M., Cohen, I., Soffer, A. & Aurbach, D. The Dependence of the Desalination Performance in Capacitive Deionization Processes on the Electrodes PZC. *J. Electrochem. Soc.* **158,** P168 (2011).

12. Doornbusch, G. J., Dykstra, J., Biesheuvel, P. M. & Suss, M. E. Fluidized bed electrodes with high carbon loading applied to water desalination by capacitive deionization. *J. Mater. Chem. A* **4,** 3642–3647 (2016).

13. Hatzell, K. B. *et al.* Effect of oxidation of carbon material on suspension electrodes for flow electrode capacitive deionization. *Environ. Sci. Technol.* **49,** 3040–3047 (2015).

14. Zhao, R., Porada, S., Biesheuvel, P. M. & Van der Wal, A. Energy consumption in membrane capacitive deionization for different water recoveries and flow rates, and comparison with reverse osmosis. *Desalination* **330,** 35–41 (2013).

15. Porada, S. *et al.* Direct prediction of the desalination performance of porous carbon electrodes for capacitive deionization. *Energy Environ. Sci.* **6,** 3700 (2013).

16. Porada, S., Zhao, R., Van Der Wal, A., Presser, V. & Biesheuvel, P. M. Review on the science and technology of water desalination by capacitive deionization. *Prog. Mater. Sci.* **58,** 1388–1442 (2013).

17. Suss, M. E. *et al.* Impedance-based study of capacitive porous carbon electrodes with hierarchical

and bimodal porosity. *J. Power Sources* **241,** 266–273 (2013).

18. Lyklema, J. & Minor, M. On surface conduction and its role in electrokinetics. **140,** 33–41 (1998).

19. Zangle, T. A., Mani, A. & Santiago, J. G. On the Propagation of Concentration Polarization from Microchannel-Nanochannel Interfaces Part II: Numerical and Experimental Study. 3909–3916 (2009).

20. Dydek, E. V. *et al.* Overlimiting Current in a Microchannel. *Phys. Rev. Lett.* **118301,** 1–5 (2011).

21. Nam, S. *et al.* Experimental Verification of Overlimiting Current by Surface Conduction and Electro-Osmotic Flow in Microchannels. *Phys. Rev. Lett.* **114501,** 1–5 (2015).

22. Mani, A., Zangle, T. A. & Santiago, J. G. On the Propagation of Concentration Polarization from Microchannel-Nanochannel Interfaces Part I: Analyical Model and Characteristic Analysis. *Langmuir* **25,** 3898–3908 (2009).

23. Bonthuis, D. J. & Netz, R. R. Unraveling the Combined Effects of Dielectric and Viscosity Profiles on Surface Capacitance, Electro-Osmotic Mobility, and Electric Surface Conductivity. *Langmuir* **28,** 16049–16059 (2012).

24. Han, J., Khoo, E., Bai, P. & Bazant, M. Z. Over-limiting Current and Control of Dendritic Growth by Surface Conduction in Nanopores. *Sci. Rep.* **4,** 1–8 (2014).

25. Deng, D. *et al.* Overlimiting Current and Shock Electrodialysis in Porous Media. *Langmuir* **29,** 16167–16177 (2013).

26. Schlumpberger, S., Lu, N. B., Suss, M. & Bazant, M. Z. Scalable and Continuous Water Deionization by Shock Electrodialysis. *Environ. Sci. Technol. Lett.* **2,** 367–372 (2015).

27. Strickland, D. G., Suss, M. E., Zangle, T. A. & Santiago, J. G. Evidence shows concentration polarization and its propagation can be key factors determining electroosmotic pump performance. *Sensors Actuators, B Chem.* **143,** 795–798 (2010).

28. Suss, M. E., Mani, A., Zangle, T. A. & Santiago, J. G. Electroosmotic pump performance is affected by concentration polarizations of both electrodes and pump. *Sensors Actuators, A Phys.* **165,** 310–315 (2011).

29. Nischang, I. *et al.* Concentration Polarization and Nonequilibrium Electroosmotic Slip in Dense Multiparticle Systems. 9271–9281 (2007).

30. Dukhin, S. S. Non-Equilibrium Electric Surface Phenomena. *Adv. Colloid Interface Sci.* **44,** 1–134 (1993).

31. Henry, D. C. The Electrophoresis of Suspended Particles. IV. The Surface Conductivity Effect. *Trans. Faraday Soc.* **44,** 1021–1026 (1948).

32. Chu, K. T. & Bazant, M. Z. Nonlinear electrochemical relaxation around conductors. *Phys. Rev. E - Stat. Nonlinear, Soft Matter Phys.* **74,** (2006).

33. Bazant, M. Z., Thornton, K. & Ajdari, A. Diffuse-charge dynamics in electrochemical systems. *Phys. Rev. E - Stat. Nonlinear, Soft Matter Phys.* **70,** 1–24 (2004).

34. Simon, P. & Gogotsi, Y. Materials for electrochemical capacitors. *Nat. Mater.* **7,** 845–854 (2008).

35. Mirzadeh, M., Gibou, F. & Squires, T. M. Enhanced charging kinetics of porous electrodes: Surface

conduction as a short-circuit mechanism. *Phys. Rev. Lett.* **113,** 1–5 (2014).

36. Vol'fkovich, Y. M., Mikhalin, A. A. & Rychagov, A. Y. Surface conductivity measurements for porous carbon electrodes. *Russ. J. Electrochem.* **49,** 594–598 (2013).

37. Biesheuvel, P. M. & Bazant, M. Z. Nonlinear dynamics of capacitive charging and desalination by porous electrodes. *Phys. Rev. E - Stat. Nonlinear, Soft Matter Phys.* **81,** 1–12 (2010).

38. Chu, K. T. & Bazant, M. Z. Surface conservation laws at microscopically diffuse interfaces. *J. Colloid Interface Sci.* **315,** 319–329 (2007).

39. Porada, S. *et al.* Water desalination using capacitive deionization with microporous carbon electrodes. *ACS Appl. Mater. Interfaces* **4,** 1194–1199 (2012).

40. Hemmatifar, A., Stadermann, M. & Santiago, J. G. Two-Dimensional Porous Electrode Model for Capacitive Deionization. *J. Phys. Chem. C* **119,** 24681–24694 (2015).

41. Biesheuvel, M., Fu, Y. & Bazant, M. Diffuse charge and Faradaic reactions in porous electrodes. *Phys. Rev. E* **83,** 061507 (2011).

42. Biesheuvel, P. M., Fu, Y. & Bazant, M. Z. Electrochemistry and capacitive charging of porous electrodes in asymmetric multicomponent electrolytes. *Russ. J. Electrochem.* **48,** 580–592 (2012).

43. Biesheuvel, P. M., Porada, S., Levi, M. & Bazant, M. Z. Attractive forces in microporous carbon electrodes for capacitive deionization. *J. Solid State Electrochem.* **18,** 1365–1376 (2014).

44. Biesheuvel, P. M. Activated carbon is an electron-conducting amphoteric ion adsorbent. *ArXiv* 1–9 (2015).

45. Newman, J. & Tiedemann, W. Porous-electrode theory with battery applications. *AIChE J.* **21,** 25–41 (1975).

46. De Levie, R. On porous electrodes in electrolyte solutions. *Electrochim. Acta* **8,** (1963).

47. Mirzadeh, M. & Gibou, F. A conservative discretization of the Poisson – Nernst – Planck equations on adaptive Cartesian grids. *J. Comput. Phys.* **274,** 633–653 (2014).